\documentclass[aps,prb,reprint,amsmath,amssymb,floatfix,citeautoscript]{revtex4-2}
\usepackage{graphicx}
\usepackage{color}
\usepackage{bm}
\usepackage{tabularx,booktabs}
\usepackage{multirow}
\usepackage[colorlinks,linkcolor=blue,citecolor=blue,urlcolor=blue]{hyperref}

\usepackage{multirow}
\providecommand{\U}[1]{\protect\rule{.1in}{.1in}}

\date{\today}

\begin{document}

\title{\textcolor{black}{Alkali doping of Zn$_{\rm x}$Mg$_{\rm 1-x}$O alloys for $p$-type conductivity}}
\author{John L. Lyons}\thanks{john.l.lyons27.civ@us.navy.mil}
\affiliation{Center for Computational Materials Science, US Naval Research Laboratory, Washington,
D.C. 20375, USA}

\begin{abstract}
Nearly all ultrawide-bandgap oxides are affected by hole localization that limits $p$-type conductivity and thus potential applications for these materials. Highly
localized holes, also known as hole polarons, trap in the vicinity of acceptor dopants, giving rise to large ionization energies and severely constraining free hole
concentrations. Though this hole-trapping behavior affects wurtzite zinc oxide, rocksalt zinc oxide was recently found to be resistant to the formation of hole polarons.
Moreover, $p$-type doping using lithium acceptors was predicted to be achievable. While rocksalt zinc oxide is metastable and has a band gap near $\sim$3 eV, here it is
found that zinc magnesium oxide (Zn$_{\rm x}$Mg$_{\rm 1-x}$O) alloys remain $p$-type dopable within the stable rocksalt crystal structure, in addition to exhibiting band
gaps in excess of 4 eV. As in rocksalt zinc oxide, alkali acceptors are shallow in zinc magnesium oxide and do not appear to be affected by donor compensation. These
results indicate that alkali-doped Zn$_{\rm x}$Mg$_{\rm 1-x}$O alloys are a promising system for achieving a $p$-type dopable ultrawide-bandgap oxide.
\end{abstract}

\maketitle

\section{Introduction}

Ultrawide-bandgap (UWBG) semiconductors are of vital concern for future power and radio frequency electronics, ultraviolet light emitters, and semiconductor devices for
extreme environments \cite{Tsao17}. Nearly all of the UWBG materials exhibit unipolar electrical conductivity, where either only $p$-type (holes) or $n$-type (electrons)
conductivity is possible, but not both. Since devices such as bipolar junction transistors and laser diodes usually require both kinds of conductivity, this shortcoming
limits the applications of existing UWBG semiconductors.

In most materials, the major difficulty has been achieving reliable $p$-type conductivity.\cite{Tsao17,Lyons24,Chae25,Vu25} (Diamond is an outlier, in that while modest
$p$-type conductivity can be achieved,\cite{Visser92} $n$-type conductivity cannot.) AlN and AlGaN have long been known to exhibit difficulties with $p$-type doping and
exhibit
very low doping efficiencies.\cite{Taniyasu06,Kneissl23} Though c-BN has shown promise\cite{Mishima87,Weston17} for acceptor doping (as well as for donor
doping\cite{Hirama20,Turiansky2021}), reports are still limited, and growth of c-BN is difficult.

Oxides, and monoclinic gallium oxide ($\beta$-Ga$_2$O$_3$) in particular, have received significant attention as UWBG semiconductors.\cite{Higashiwaki16,Tsao17,Tadjer19} However, holes have a tendency to strongly self trap in this class of materials.\cite{varley12,Lyons22,Vu25} These self-trapped holes (or hole polarons) strongly bind to acceptor dopants
and give rise to large acceptor ionization energies.\cite{Kyrtsos18,Lyons18} Recent theoretical studies indicate that rutile germanium oxide (r-GeO$_2$)\cite{Chae19} and
rutile silicon dioxide (r-SiO$_2$)\cite{Lyons23} might be less susceptible to hole localization, potentially enabling $p$-type conductivity in the rutile oxide materials
family. However, methods to synthesize r-GeO$_2$ are still in their infancy,\cite{Niedermeier20,Rahaman24,Galazka24} and r-SiO$_2$ is a highly metastable phase\cite{Linn18}
which will impair growth efforts. Other UWBG oxides have been proposed to potentially exhibit $p$-type conductivity, such as LiGa$_5$O$_8$,\cite{Zhang25} but calculations
indicate this material is also susceptible to hole localization and deep acceptor behavior.\cite{Lyons24a,Dabsamut24}

Rocksalt oxides are a much less touted materials system among UWBG semiconductors, in comparison to $\beta$-Ga$_2$O$_3$ or r-GeO$_2$. Much less is known about the behavior of
defects and dopants in these systems, despite the prevalence of applications for MgO.\cite{Rinke12} While dopants such as Li have experimentally been shown to be relatively
deep acceptors in MgO\cite{Tardio02}, the reported ionization energies (near 0.7 eV) are smaller than other UWBG oxides such as Ga$_2$O$_3$. Intriguingly, Goyal and
Stevanovi\'{c} used hybrid functional calculations to show\cite{Goyal18} that rocksalt zinc oxide (rs-ZnO), a metastable polymorph, was $p$-type dopable. Though the more
stable wurtzite zinc oxide (wz-ZnO) also suffers from hole trapping and deep acceptor dopants\cite{Lyons09}, rs-ZnO has a higher valence-band edge and lower hole effective
mass than wz-ZnO, both of which destabilize hole localization.

Though growth of rs-ZnO is likely difficult\cite{Decremps02}, rocksalt is the most stable crystal structure of magnesium oxide. A wide range of rs-Zn$_{\rm x}$Mg$_{\rm 1-x}$O
alloys have been grown via pulsed laser deposition\cite{Choopun02} or molecular beam epitaxy.\cite{Lu16,Gorczyca20} Moreover, such alloys might be expected to exhibit the
high densities that are thought to promote hole delocalization in UWBG oxides.\cite{Lyons24,Chae25} Taken together, these results suggest that rs-Zn$_{\rm x}$Mg$_{\rm 1-x}$O
might represent a readily synthesizable, $p$-type dopable UWBG oxide.

In this study, this possibility is explored using hybrid density functional theory. In agreement with prior experiments, alkali acceptors in rs-MgO are found to have high
ionization energies, as they are affected by hole localization. However, calculations of the electronic structure of Zn$_{\rm x}$Mg$_{\rm 1-x}$O alloys indicate that they
share similar properties to rs-ZnO. Explicit defect modeling in the Zn$_{\rm x}$Mg$_{\rm 1-x}$O alloys (for both ordered and disordered structures) show that alkali acceptors
are shallow dopants in these alloys, and that under O-rich conditions compensation of $p$-type doping can be avoided.

\section{Methods}

To calculate defects and dopants in MgO and Zn$_{\rm x}$Mg$_{\rm 1-x}$O alloys, hybrid DFT calculations are employed using the projector-augmented wave method~\cite{blochl94}
implemented in VASP~\cite{kresse99}, and with a 400 eV planewave cutoff. 8$\times$8$\times$8 $\Gamma$-centered $k$-point meshes were used to optimize the lattice parameters
of primitive cells of rs-ZnO and rs-MgO, and 6$\times$6$\times$6 $\Gamma$-centered $k$-point meshes were used to optimize the lattice parameters of 8-atom cells of ordered
Zn$_{\rm x}$Mg$_{\rm 1-x}$O alloys.

The HSE hybrid functional~\cite{HSE,HSEe} with 34.5\% mixing and a screening parameter of 0.2 \AA$^{-1}$
is used for all compounds in this work. This provides an excellent description of the band gap\cite{Whited69} and lattice parameters\cite{Hirata77} for MgO, which are in good
agreement with experimentally observed values and prior theoretical work\cite{Schleife06,Rinke12}. The HSE-calculated structural and electronic properties of ZnO, MgO, and
the ordered
Zn$_{\rm 0.5}$Mg$_{\rm 0.5}$O alloy, all in the rocksalt crystal structure, are listed in Table~1.

\begin{table}
\begin{tabular}{lcccc}
  \hline
  \multirow{3}{*}{\textbf{material}} &
  \multicolumn{2}{c}{\textbf{bandgap (eV)}} &
  \multicolumn{2}{c}{\textbf{$a$ (${\rm \AA}$)}} \\
  & calc. & exp. & calc. & exp.\\ \hline
  \toprule
  rs-ZnO & 3.57 ($i$) & 2.75\cite{Sans05} & 4.24 & 4.27\cite{Karzel96}~\\
  rs-Zn$_{\rm 0.5}$Mg$_{\rm 0.5}$O & 4.41 ($i$) & -- & $4.20$ & --\\
  rs-MgO & 7.55 ($d$) & 7.78\cite{Whited69} & 4.16 & 4.22\cite{Hirata77}~\\
\end{tabular}
\caption{Structure and electronic properties of ZnO, MgO, and the ordered Zn$_{\rm 0.5}$Mg$_{\rm 0.5}$O alloy in the rocksalt crystal structure. Band gaps are listed in eV,
and lattice parameters are in \AA. ZnO and Zn$_{\rm 0.5}$Mg$_{\rm 0.5}$O have indirect ($i$) bandgaps, while MgO has a direct ($d$) bandgap at $\Gamma$; full band structures
are provided in Appendix A.}
\label{TAB:bulk}
\end{table}

For defect calculations, 216-atom supercells with one $k$-point at $\Gamma$ are used for both MgO and the ordered-alloy calculations. (For the random alloy structures, which
are discussed in Appendix B, 64-atom rocksalt supercells are used.) The formalism established in Ref.~\onlinecite{FreysoldtRMP} is used to calculate defect formation
energies. Taking the example of the lithium substitutional acceptor in MgO (Li$_{\rm Mg}$) as an example, its formation energy in charge state $q$ is given by:
\begin{equation}
\label{eqeform}
\begin{aligned}
E^f({\rm Li}_{\rm Mg}^q)= E_{\rm tot}({\rm Li}_{\rm Mg}^q) - E_{\rm tot}({\rm MgO}) + \mu_{\rm Mg} - \mu_{\rm Li}\\
 + q (\varepsilon_{F} + \varepsilon_{v}) + \Delta^q, 
\end{aligned}
\end{equation}
where $E_{\rm tot}({\rm Li}_{\rm Mg}^q)$ is the total energy of a supercell containing the Li acceptor in charge state $q$, and $E_{\rm tot}({\rm MgO})$ is the total energy
of pristine MgO using the same supercell. Any electrons added to or removed from the supercell are exchanged with the electron reservoir in the semiconductor host, i.e., the
Fermi level ($\varepsilon_{F}$), which is referenced to the VBM of MgO ($\varepsilon_v$). The final term in Eq.~\ref{eqeform}, $\Delta^q$, is the charge-state dependent term
that corrects for the finite size of the charged supercell \cite{defectschemePRL,Freysoldt2011}.

The removed Mg atom is placed in a reservoir of energy $\mu_{\rm Mg}$, referenced to the energy per atom of Mg metal. The range of $\mu_{\rm Mg}$ is limited by the enthalpy
of formation of MgO [$\Delta {\rm H}_f ({\rm MgO})$, calculated to be $-$5.80 eV], so $\mu_{\rm Mg}$ can take values between 0 eV (the Mg-rich limit) and $-5.80$ eV (the
O-rich limit). Analogous considerations apply to $\mu_{\rm O}$, which is referenced to one half of the energy of the spin-polarized O$_2$ molecule.

The chemical potential for the impurity is limited by the formation of secondary phases; in the case of Li, $\mu_{\rm Li}$ is limited by the formation of Li$_2$O through the
relation:
\begin{equation}
\label{alform}
2\mu_{\rm Li} + \mu_{\rm O} < \Delta {\rm H}_f({\rm Li}_2{\rm O}),
\end{equation}
where $\Delta$H$_f$(Li$_2$O) is the formation enthalpy of Li$_2$O [calculated to be $-5.85$ eV]. Analogous expressions can be written for the secondary phases that limit the
chemical potentials of other impurities (e.g., Na$_2$O for Na and K$_2$O for K).

Band alignments between the binary compounds and Zn$_{\rm x}$Mg$_{\rm 1-x}$O ordered alloys are calculated using surface-slab calculations, which align the VBM of each oxide (referenced to the averaged electrostatic potential within each compound) to the vacuum level. This approach provides a common reference and avoids strain effects
\cite{Franciosi96}. Nonpolar (100) surfaces are used for these calculations, with eight bilayers of oxide (at least 15~\AA~of material) together with 25~\AA~of vacuum. Atoms
within 5~\AA~of each surface were relaxed, while interior atoms had positions fixed in order to mimic bulk material.

\section{Results and discussion}

\subsection{Bulk properties of rocksalt ZnO, Zn$_{\rm 0.5}$Mg$_{\rm 0.5}$O, and MgO}
The HSE-calculated band gaps and lattice parameters for rocksalt MgO, ZnO, and Zn$_{\rm 0.5}$Mg$_{\rm 0.5}$O are compared in Table~\ref{TAB:bulk} against the available
experimental values. Both the calculated lattice parameter (4.16~\AA) and band gap (7.55 eV) of MgO are in good agreement with experiments, which yielded values of
4.22~\AA\cite{Hirata77} and 7.78 eV\cite{Whited69}. The calculated lattice parameter of rs-ZnO (4.24~\AA) is also in good agreement with the experimental value
(4.27~\AA)\cite{Karzel96}. The lattice parameter of Zn$_{\rm 0.5}$Mg$_{\rm 0.5}$O falls halfway between MgO and ZnO, as would be expected by Vegard's law.

While the calculated band gap of rs-ZnO (3.57 eV) is larger than the reported experimental gap of 2.75 eV\cite{Sans05}, reports on this material are limited. Moreover, due to
its metastability, crystal quality might limit reliability. It should also be noted that the higher gap reported here is more consistent with prior theoretical
reports\cite{Goyal18,Chae25}, and that similarly high mixing parameters are needed to accurately describe the electronic structure of wz-ZnO.\cite{Lyons09}

In agreement with previous work,\cite{Schleife06} MgO has a direct band gap at $\Gamma$; while the indirect gap of rs-ZnO (with the VBM at L and the conduction-band minimum
at $\Gamma$) also agrees with the results from Ref.~\onlinecite{Goyal18}. The band gap of ordered Zn$_{\rm 0.5}$Mg$_{\rm 0.5}$O (which is also indirect) deviates from
Vegard's law, but is consistent with the band-gap bowing that has previously been observed\cite{Gorczyca20} in this alloy system. HSE-calculated band structures are included
in Fig.~\ref{fig:BS} of Appendix A.

\subsection{Doping of rocksalt MgO}
\begin{figure}
\includegraphics[width=\columnwidth]{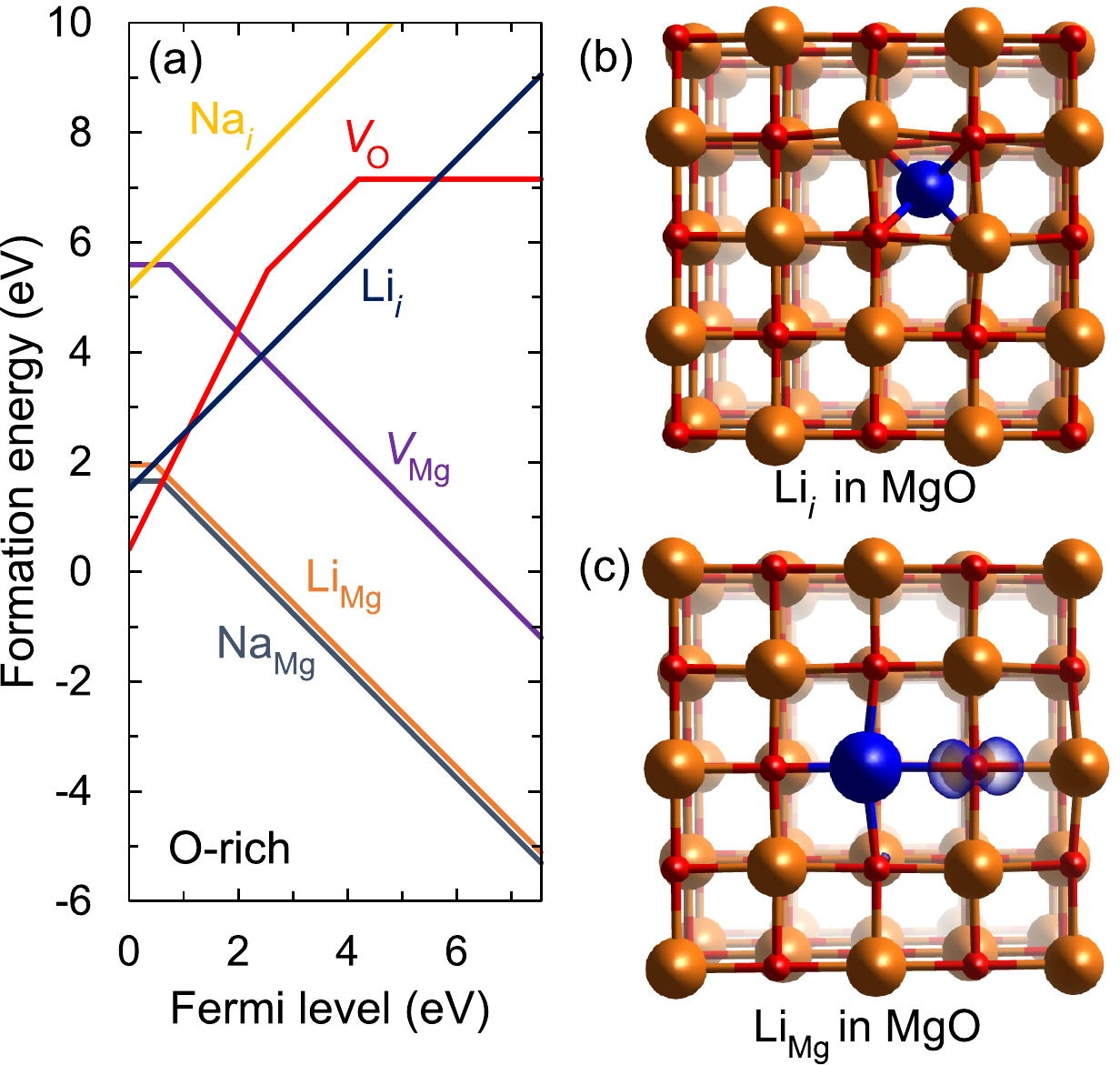}
\caption{\label{fig:MgO} (a) Formation energy versus Fermi level for impurities and defects in MgO under O-rich conditions. (b) Structure of the Li$_{i}$ donor in MgO, with
Mg atoms in orange, O atoms in red, and Li in blue. (c) Isosurface (in blue) of the spin density for the Li$_{\rm Mg}$ acceptor in MgO. (Plotted using
VESTA\protect\cite{VESTA}).}
\end{figure}

In Fig.~\ref{fig:MgO} the results for alkali doping of rs-MgO are summarized, showing that Li and Na can incorporate as acceptors on the Mg site. Unlike in rs-ZnO, alkali
substitutionals can trap hole polarons in rs-MgO and exhibit deep transition levels. As shown in Fig.~\ref{fig:MgO}c, these holes trap onto an oxygen site that is a nearest
neighbor to the substitutional impurity. This hole localization results in large acceptor ionization energies: the smallest occurs for Na$_{\rm Mg}$ (0.49 eV), while those of
Li$_{\rm Mg}$ (0.59 eV) and K$_{\rm Mg}$ (0.76 eV) are larger. The calculated transition level of Li$_{\rm Mg}$ (0.59 eV) is similar to a prior experimental
report\cite{Tardio02} of a 0.7 eV activation energy measured in Li-doped MgO.

These results indicate that rs-MgO will itself not be highly $p$-type dopable with alkali elements, as the large ionization energies will limit the number of free hole
concentrations that can arise from acceptor doping. Still, the ionization energies calculated here for rs-MgO are only slightly larger than those of GeO$_2$\cite{Chae19}, and
are smaller than many other UWBG oxides\cite{Kyrtsos18,Lyons18,Lyons22}.

The formation energies of possible compensating species, including oxygen vacancies ($V_{\rm O}$) and alkali donor interstitials (including Na$_i$ and Li$_i$), are also shown
in Fig.~\ref{fig:MgO}. The O-rich extreme is considered; as discussed in Ref.~\onlinecite{Goyal18}, these conditions are most favorable for the incorporation of the alkali
substitutional acceptors. For most Fermi levels, Li$_{\rm Mg}$ and Na$_{\rm Mg}$ acceptors (which have nearly the same formation energy) are the lowest-energy defects among
those considered. Compensating self interstitials (e.g., Na$_i$ and Li$_i$) are much higher in energy over most of the band gap; only when $\varepsilon_{F}$ nears 0 eV (i.e.,
$p$-type conditions) does the formation energy of Li$_i$ approach that of Li$_{\rm Mg}$. In contrast, Na$_i$ is 3 eV higher in energy than Na$_{\rm Mg}$, even when
$\varepsilon_{F}$ = 0 eV.

These results suggest that, like in rs-ZnO, alkali acceptor dopants in MgO can avoid being compensated by native defects or alkali interstitial donors. Thus, if the
ionization
energies of the acceptor dopants could somehow be lowered, $p$-type conductivity could be achieved in this material family. Because alkali acceptors in rs-ZnO have been
previously found to exhibit small ionization energies, this suggests that Zn$_{\rm x}$Mg$_{\rm 1-x}$O alloys might be a viable system for achieving $p$-type conductivity.
Before exploring this possibility, the role of Zn alloying on the rs-MgO electronic structure is first considered.

\subsection{Electronic structure of ordered Zn$_{\rm x}$Mg$_{\rm 1-x}$O alloys}
Incorporating Mg into rs-ZnO increases the band gap of the Zn$_{\rm x}$Mg$_{\rm 1-x}$O alloy, but it does not substantively affect other aspects of the band structure, as
shown in Fig.~\ref{fig:BS}b-d of Appendix A. The VBM remains at the L point for the intermediate alloys, as the relative energy of the conduction-band minimum (CBM) at $\Gamma$ moves
upward. This yields indirect bandgaps of 3.88 eV for Zn$_{\rm 0.75}$Mg$_{\rm 0.25}$O, 4.40 eV for Zn$_{\rm 0.5}$Mg$_{\rm 0.5}$O, and 5.16 eV for Zn$_{\rm 0.25}$Mg$_{\rm
0.75}$O.

Note that these values are for the case of ordered alloys. For the random alloy, Zn$_{\rm 0.5}$Mg$_{\rm 0.5}$O is calculated to have a band gap only 0.16 eV larger (4.56 eV),
suggesting that disorder does not have a major effect on the band structure of the oxide. It should also be noted that disorder does not strongly affect the energetics of the
Zn$_{\rm x}$Mg$_{\rm 1-x}$O alloy, as the disordered compound is within 0.01 eV/formula unit of the ordered structure for Zn$_{\rm 0.5}$Mg$_{\rm 0.5}$O.

\begin{figure}
\includegraphics[width=\columnwidth]{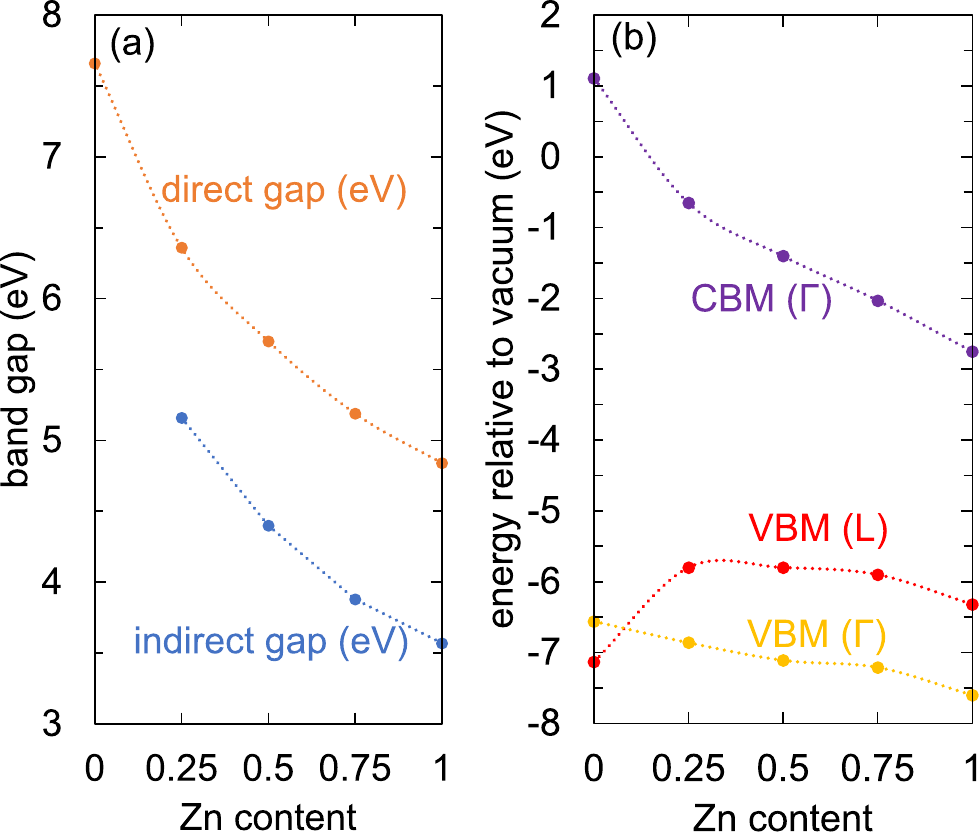}
\caption{\label{fig:trends} (a) Direct and indirect band gaps as a function of Zn content in Zn$_{\rm x}$Mg$_{\rm 1-x}$O alloys. (b) Energies (in eV, with respect to
vacuum) for the different high-symmetry points in the band structures of Zn$_{\rm x}$Mg$_{\rm 1-x}$O alloys. Dotted lines are guides for the eye.}
\end{figure}

The trends in the electronic structure can be seen in Fig.~\ref{fig:trends}. As shown in Fig.~\ref{fig:trends}a, the band gap remains indirect in rs-ZnO and in the Zn$_{\rm
x}$Mg$_{\rm 1-x}$O alloys. Both the direct and indirect gaps increase rapidly as Mg content increases. In MgO, the valence band at the L point moves below the VBM at $\Gamma$,
leaving a large direct gap (as shown in Appendix A).

In Fig.~\ref{fig:trends}b the absolute trends in energies of the bands can be observed for the ordered alloys. (As discussed in the Methods section, these energies are referenced to the vacuum level via surface-slab calculations.) These results show that as the Mg content in the Zn$_{\rm x}$Mg$_{\rm 1-x}$O alloy is increased, the L-point VBM rises slightly in energy, while the CBM (at $\Gamma$) increases rapidly with Mg content. Only once Zn is no longer present (i.e., for pure MgO) does the VBM revert to the $\Gamma$ point, yielding the large direct gap of MgO. These results suggest that the electronic structure of rs-Zn$_{\rm x}$Mg$_{\rm 1-x}$O alloys will be similar to that of rs-ZnO, with the valence-band properties being driven by the maximum at the L point (albeit with a larger band gap). Thus, if the ionization energies in the alloy are similar to rs-ZnO, $p$-type dopability should also be achieved.

\subsection{Doping of rocksalt Zn$_{\rm 0.5}$Mg$_{\rm 0.5}$O}
\begin{figure}
\includegraphics[width=\columnwidth]{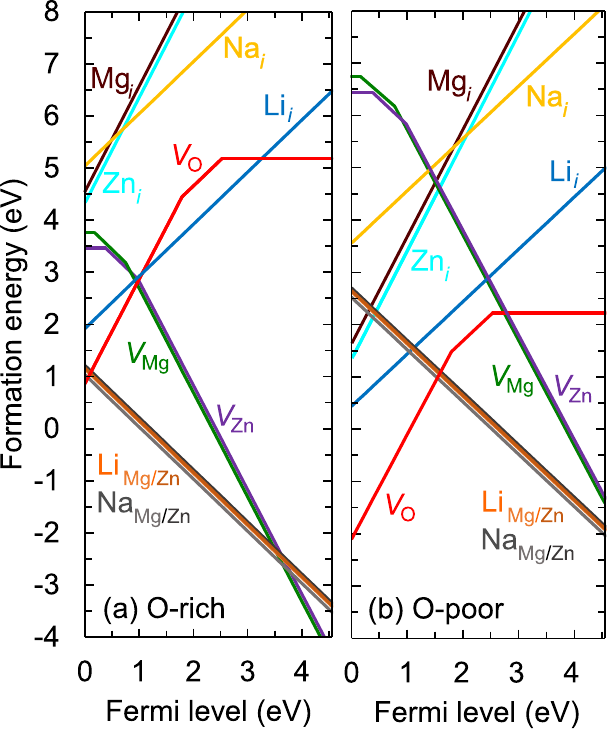}
\caption{\label{fig:order} Formation energy versus Fermi level for impurities and defects in Zn$_{\rm 0.5}$Mg$_{\rm 0.5}$O under (a) O-rich and (b) O-poor conditions. The
formation energies of the
alkali substitutional acceptor dopants (Li$_{\rm Zn}$ in dark orange, Li$_{\rm Mg}$ in light orange, Na$_{\rm Zn}$ in light gray, and Na$_{\rm Mg}$ in dark grey) are
clustered at similar energies in both cases.}
\end{figure}

The behavior of alkali impurities in Zn$_{\rm 0.5}$Mg$_{\rm 0.5}$O is found to be similar to what was reported previously in Ref.~\onlinecite{Goyal18} for rs-ZnO. In
Fig.~\ref{fig:order}, the formation energy (in eV) versus Fermi level (also in eV) is plotted for alkali substitutional acceptors and compared with native defects and alkali interstititals. (Though these calculations used an ordered Zn$_{\rm 0.5}$Mg$_{\rm 0.5}$O supercell, the results presented in Appendix B indicate that disorder will not strongly affect the conclusions.) As with rs-ZnO,\cite{Goyal18} the alkali substitutional acceptors are shallow, and do not exhibit deep transition levels. This behavior is attributed to the high-lying L-point VBM, which suppresses hole polaron formation, as was also reported for rs-ZnO. Both Li and Na substitutional acceptors are found to act as effective-mass acceptors, with ionization energies near $\sim$0.1-0.2 eV.

The energy of the Li and Na cation-site substitutional acceptors are quite similar, regardless of which site onto which they incorporate. As seen in Fig.~\ref{fig:order}, the
formation energies of Li$_{\rm Zn}$, Li$_{\rm Mg}$, Na$_{\rm Zn}$, and Na$_{\rm Mg}$ cluster within about 0.2 eV. This can be attributed to the fact that both Zn and Mg are
surrounded only by O nearest neighbors in the Zn$_{\rm x}$Mg$_{\rm 1-x}$O alloy; since defect properties are driven mainly by nearest neighbors, it follows that cation-site
substitutionals from similar impurities will lead to similar properties. (Similar conclusions were reached in a prior study of substitutional dopants in InGaN
alloys.\cite{Wickramaratne20}) Choice of cation site also minimally affects the structural properties: Na$_{\rm Zn}^-$ features Na$-$O average bond lengths of 2.18~\AA~versus
2.21~\AA~for Na$_{\rm Mg}^{-}$, while Li$_{\rm Zn}^{-}$ has Li$-$O bond lengths of 2.12~\AA~versus 2.16~\AA~for Li$_{\rm Mg}^{-}$. This behavior is not likely to extend to
the larger alkali elements, however: potassium substituting for Zn or Mg in ordered Zn$_{\rm 0.5}$Mg$_{\rm 0.5}$O leads to polaronic acceptor behavior, with ionization
energies near 0.3 eV.

Competition with native acceptors will not apparently affect the incorporation of alkali dopants. Under $n$-type conditions (e.g., when the Fermi level is near the CBM), the cation vacancies ($V_{\rm Zn}$ and $V_{\rm Mg}$) have lower formation energies than the alkali substitutional acceptors under O-rich conditions. However, as the Fermi level moves towards the VBM (e.g., under $p$-type conditions), the alkali substitutionals become lower in energy.

Compensation by donors also appears to be avoidable under the O-rich conditions shown in Fig.~\ref{fig:order}a. Both alkali interstitial donors are significantly higher in
energy than the alkali substitutional acceptors, especially under O-rich conditions, and neither should be a source of compensation in alkali-doped material. Only when the
Fermi level is at the VBM does the formation of $V_{\rm O}$ approach that of the acceptor dopants, suggesting that these species will also not be a source of compensation
under O-rich conditions. Native interstitial donors (e.g., Zn$_i$ and Mg$_i$) have higher formation energies than Li$_i$ under all conditions.

Based on these results, sodium seems to be the best candidate for acceptor doping of Zn$_{\rm 0.5}$Mg$_{\rm 0.5}$O, since its donor interstitial is high in energy and does
not appear to be a potential compensating species. Oxygen vacancies do appear to be the most likely compensating donors, but their presence could be minimized with growth
under the O-rich extreme. Prior work\cite{Tardio02} on Li doping of MgO suggested that oxidation of doped crystals was important for electrically activating the acceptor; similar approaches might also be explored for the Zn$_{\rm x}$Mg$_{\rm 1-x}$O alloys.

\section{Summary and conclusions}
Using hybrid density functional calculations, the electronic structures of Zn$_{\rm x}$Mg$_{\rm 1-x}$O alloys were calculated, and the behavior of alkali dopants was
determined in both rocksalt MgO and Zn$_{\rm 0.5}$Mg$_{\rm 0.5}$O. Like rocksalt ZnO, the Zn$_{\rm x}$Mg$_{\rm 1-x}$O alloys exhibit indirect band gaps, with the valence-band
maxima occurring away from the $\Gamma$ point. This higher-lying band promotes hole delocalization, and leads to shallow acceptor behavior for Li and Na substitutional
acceptors. Moreover, acceptor doping should be free from compensation, at least towards O-rich conditions, as these acceptor impurities have formation energies lower than potential compensating donors such as oxygen vacancies and alkali interstitials. As the synthesis of rocksalt Zn$_{\rm x}$Mg$_{\rm 1-x}$O alloys has already proven feasible, this study indicates that these materials are a promising direction for $p$-type-dopable ultrawide-bandgap oxides.

\section{Acknowledgements}

This work was supported by the Office of Naval Research through the Naval Research Laboratory's Base Research Program. Computations were performed at the DoD Major Shared
Resource Centers at AFRL and the Army ERDC. Henryk Teisseyre is thanked for illuminating discussions.

%
\section{Data Availability}
The data the support the findings of this study are available upon request.

\begin{figure*}
\includegraphics[width=\textwidth]{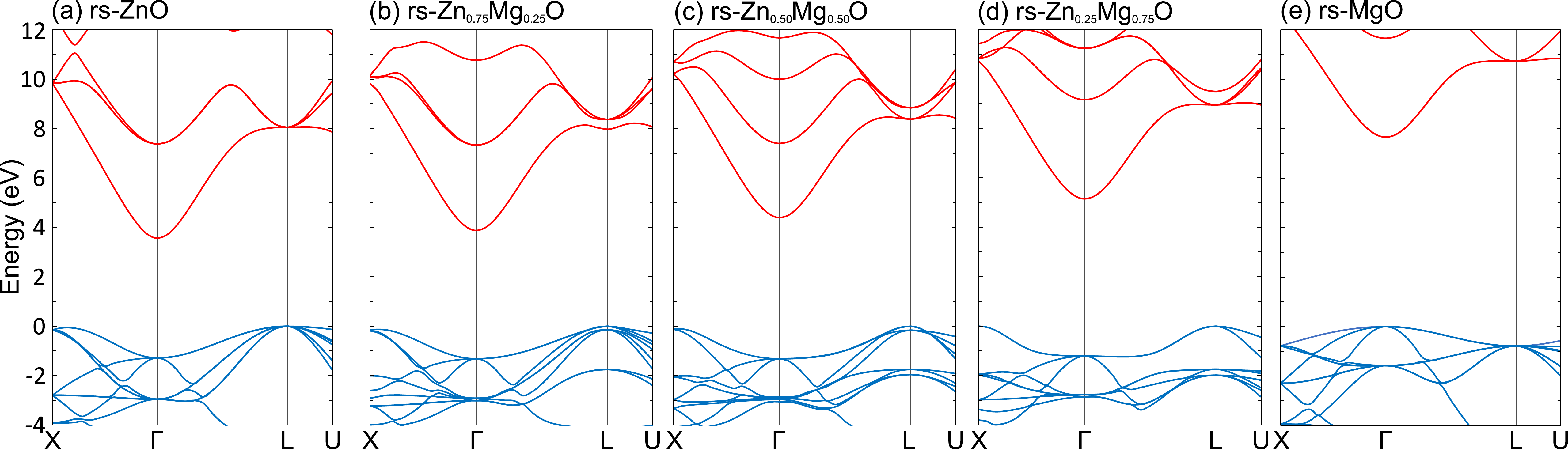}
\caption{\label{fig:BS} HSE-calculated band structures of (a) rs-ZnO, (b) rs-Zn$_{\rm 0.75}$Mg$_{\rm 0.25}$O, (c) rs-Zn$_{\rm 0.5}$Mg$_{\rm 0.5}$O, (d) rs-Zn$_{\rm
0.25}$Mg$_{\rm 0.75}$O, and (e) rs-MgO. Valence bands are drawn in blue, while conduction bands are drawn in red, and the VBM is set to 0 eV for each system.}
\end{figure*}

\section{Appendix A: Band structures of compounds}
The band structures of ZnO, MgO, and ordered Zn$_{\rm x}$Mg$_{\rm 1-x}$O alloys, all in the rocksalt crystal structure, are shown in Fig.~\ref{fig:BS}. All calculations
utilize the self-consistent HSE hybrid functional, with 34.5\% exact exchange and a screening parameter of 0.2~\AA$^{-1}$. As in prior studies~\cite{Lany14,Goyal18}, it is
found that rs-ZnO has an indirect band gap, with the VBM located at the L point and the CBM located at the $\Gamma$ point. The calculated rs-ZnO band gap is 3.57 eV, as shown
in Table~\ref{TAB:bulk}. Due to the use of a higher mixing parameter, the calculated band gap is larger than in prior studies, as discussed in the main text.

Consistent with prior reports,\cite{Schleife06} rs-MgO has a direct band gap at $\Gamma$, along with a much larger band gap (7.55 eV) than rs-ZnO. The calculated band gap of
rs-MgO is close to the experimental band gap of 7.78 eV\cite{Whited69}, and is similar to results from $GW$
calculations.\cite{Rinke12}

As Mg increases in the Zn$_{\rm x}$Mg$_{\rm 1-x}$O alloy, the band gaps steadily increase, with noticeable downward band bowing. This behavior is consistent with prior
studies,\cite{Gorczyca20} though the calculated band gaps fall at the lower end of the reported values. This may be due in part to the use of ordered alloys in this work: the
disordered alloy of Zn$_{\rm 0.5}$Mg$_{\rm 0.5}$O is calculated here to have a band gap 0.15 eV larger than the ordered structure.

For all cases, the VBM is set to 0 eV (i.e., the band structures are not aligned to a common reference). However, Fig.~\ref{fig:trends}b in the main text displays how
band-edge energies change as a function of energy.

\begin{figure}
\includegraphics[width=\columnwidth]{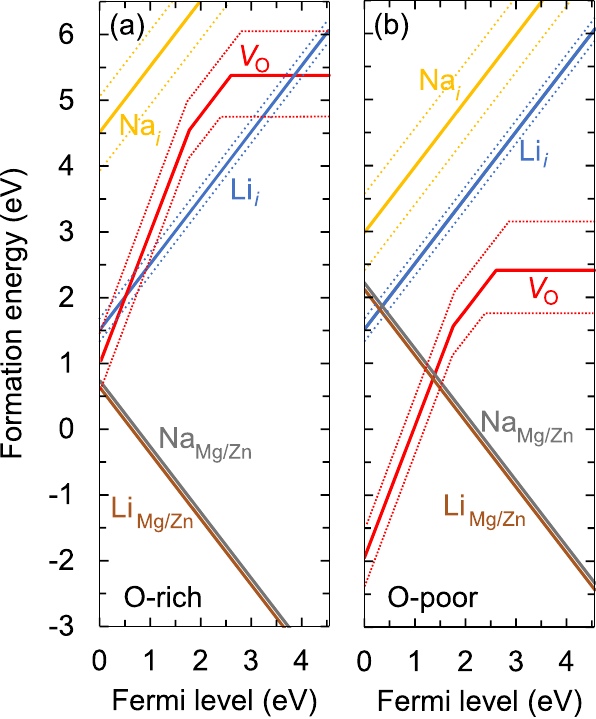}
\caption{\label{fig:sqs} Formation energy versus Fermi level for impurities and defects in rs-Zn$_{\rm 0.5}$Mg$_{\rm 0.5}$O under (a) O-rich and (b) O-poor conditions.
Solid lines represent the formation energies averaged over all configurations, while dotted lines represent a standard deviation on each side of the average.}
\end{figure}

\section{Appendix B: Defect formation energies in random alloys}

The formation energies of defects and dopants versus the Fermi level are shown in Fig.~\ref{fig:sqs} for the case of rs-Zn$_{\rm 0.5}$Mg$_{\rm 0.5}$O in a disordered 64-atom
supercell. For these calculations, six different configurations for each defect are selected (at distinct sites in the supercell), and formation energies are averaged over
the resulting formation energies. Solid lines in the figure indicate the average of these formation energies, while the color-coded dotted lines indicate the standard
deviation of formation energies.

In agreement with the results discussed in Fig.~\ref{fig:order} of the main text, there is little variability among the cation-site substitutional acceptors (i.e., Na$_{\rm
Zn}$, Na$_{\rm Mg}$, Li$_{\rm Zn}$, and Li$_{\rm Mg}$). Again, this can be attributed to the fact that these acceptors have only oxygen atoms as nearest neighbors. Thus their
character, which is mainly driven by nearest neighbors,\cite{Wickramaratne20} is not strongly impacted by Zn/Mg disorder. In agreement with the ordered-cell calculations,
acceptor dopants in the disordered cell are most stable under O-rich conditions, where they do not appear to be strongly affected by compensating donors (such as $V_{\rm
O}$ and Li$_i$/Na$_i$). All acceptor dopants also act as shallow acceptors, in agreement with the ordered-supercell calculations.

Larger variability is observed for the formation energies of the compensating donors Na$_i$, Li$_i$, and $V_{\rm O}$. The range of formation energies is largest for $V_{\rm
O}^0$ (nearly 1.5 eV) and for Na$_i$ (1.2 eV). Unlike the substitutional alkali acceptors, these species can be affected by alloy disorder, because their nearest neighbors are
not necessarily cations.

Nevertheless, the conclusions that can be drawn from the disordered-supercell results of Fig.~\ref{fig:sqs} are similar to those from Fig.~\ref{fig:order} in the main text.
Namely, that acceptor doping should be most affective under O-rich conditions, where the alkali interstitials (Na$_i$ and Li$_i$) are high in energy, and that $V_{\rm O}$ only
compete with the dopants when the Fermi level is near the VBM. As was the case for the ordered-supercell calculations, the formation energies of the acceptor dopants are
higher in O-poor conditions, where $V_{\rm O}$ becomes the dominant defect when the Fermi level moves to within 1.5 eV of the VBM.

\section{References}

%

\end{document}